\documentclass[a4paper,11pt]{article}
\usepackage{fullpage}
\usepackage{times}
\usepackage{subfigure}
\usepackage{vmargin}
\usepackage{fancyheadings}
\usepackage[cp1251]{inputenc}
\usepackage[english]{babel}
\usepackage{graphicx}
\usepackage{color}
\usepackage{cite}
\usepackage{amsmath,amssymb,amsthm,enumerate,bbm,color,verbatim,setspace}

\usepackage{braket}
\theoremstyle{plain}  
\newtheorem{thm}{Theorem}[section]

\newcommand{\rank}{\text{rank}}
\newtheorem{exm}{Example}[section]
\newtheorem{theorem}{Theorem}[section]
\newtheorem{lemma}{Lemma}[section]
\theoremstyle{remark}

\date{}
\title{Phase shift rule with the optimal parameter selection}
\author{L.A. Markovich$^{1*}$, S. Malikis$^{1}$, S.  Polla $^{1}$ and J.T. Brugu\'es$^{1,2}$\\
      $^1$Instituut-Lorentz, Universiteit Leiden, P.O. Box 9506,\\ 2300 RA Leiden, The Netherlands\\
 $^2$  Applied Quantum Algorithms Leiden, The Netherlands\\
$^*$Corresponding author e-mail: markovich@mail.lorentz.leidenuniv.nl}
\date{}
\begin{document}
\maketitle
\pagenumbering{arabic}
\begin{abstract}\noindent

The phase shift rules enable the estimation of the derivative of a quantum state with respect to phase parameters, providing valuable insights into the behavior and dynamics of quantum systems. This capability is essential in quantum simulation tasks where understanding the behavior of complex quantum systems is of interest, such as simulating chemical reactions or condensed matter systems. However, parameter shift rules are typically designed for Hamiltonian systems with equidistant eigenvalues. For systems with closely spaced eigenvalues, effective rules have not been established. We provide insights about the optimal design of a parameter shift rule, tailored to various sorts of spectral information that may be available.
The proposed method lets derivatives be calculated for any system, regardless of how close the eigenvalues are to each other. It also optimizes the number of phase shifts, which reduces the amount of gate resources needed.
\end{abstract}
\medskip

\section{Introduction}
\par Many near-term quantum computing methods are based on Variational Circuits \cite{peruzzo2014variational,schuld2020circuit}, sequences of quantum gates tuned recursively for addressing specific tasks based on classical parameters.For example, the Quantum Approximate Optimization Algorithm (QAOA) \cite{farhi2014quantum} and Variational Quantum Eigensolver (VQE) \cite{peruzzo2014variational,kandala2017hardware} heavily rely on derivative estimation to guide the parameter optimization process, leading to improved efficiency and better outcomes. Beyond optimization, derivative estimation finds its importance in scientific and engineering fields, where solving differential equations and numerical integration are paramount. Accurate knowledge of derivatives enables precise modeling and simulation of complex systems.Moreover, quantum machine learning algorithms, such as quantum neural networks or quantum support vector machines, strongly rely on derivative estimations and enhance learning capabilities.
\par Since the output of a variational quantum circuit provides a probabilistic result, the expectation value of an observable is considered an estimate of the variable.The mean values of the simple variables can be determined by taking the average over measurement results, but finding the expectation value of multi-qubit observables is more complicated and can be done by different approaches involving the quantum phase estimation algorithm~\cite{Kitaev1995,Kimmel2015,Berg2020,Wiebe2016,O_Brien2019}, quantum energy (expectation) estimation method of decomposing the observable into a weighted sum of multi-qubit Pauli strings~\cite{Peruzzo2014}, some intermediary approaches between both\cite{Wang2019,Hamamura2020,Crawford2021} or using the recently introduced single qubit quantum memory approach \cite{https://doi.org/10.48550/arxiv.2212.07710}.\par Therefore, there is a desire to formally define the gradient as a derivative of these averages.with respect to the variational parameters of the circuit. In literature, one can find different mathematical ways of calculating the underlined derivatives, like simply taking the finite difference methods or the more advanced robust polynomial interpolation technique \cite{kane_robust_2017,franca2022efficient}.However, if we are talking about the exact calculation of the derivative, this definition is difficult to implement on hardware. The reason is that we can't really "take the derivative from the gates" that realize the necessary quantum hardware because such mathematical objects can't be unitary and hence can't be realized like quantum gates. That brings us to the need for realization of the derivative using combinations of quantum implementable operations.
\par Parameter-shift rules (PSRs) are the recipes for how to get partial derivatives by evaluating parameter-shifted instances of a variational circuit. They were originally introduced to quantum machine learning in \cite{PhysRevA.98.032309,PhysRevA.99.032331}.The PSRs relate the gradient of the mean value $f$ by some parameter $t$ to evaluations of the function itself at different points:\begin{equation}\label{0} \frac{\partial f(t)}{\partial t} = \sum_{x=1}^{m} b_x (\vec{\phi}) f(t+ \phi_x),\end{equation}where the $m$-vector of the shift parameters is $\vec{\phi}=\{\phi_x\}_{x=1}^{m}$, and  $b (\vec{\phi})=\{b_x (\vec{\phi})\}_{x=1}^{m}$ is a vector of coefficients. The original two-term PSR is based on the gates with two distinct eigenvalues \cite{li2017hybrid,mitarai2018quantum,schuld2019evaluating}.  Different variations of the original PSRs can be found in literature\cite{PhysRevA.103.012405,anselmetti2021local} preserving the restriction on the amount of the Hamiltonians eigenvalues. In \cite{banchi2021measuring} the stochastic parameter-shift rule is introduced that, in combination with the generalized shift rule \cite{Wierichs2022generalparameter} allows for the differentiation of any unitary with equidistant frequencies. The strong point of these rules is the unbiased estimate of the derivative without any additional hardware costs.
\par However, we point out that the latter rules are restricted to the evenly spaced phase shifts, and no attention is paid to Hamilton's eigenvalue structure. For example, in the case of non-equidistant eigenvalues of the Hamiltonian, the latter can be close to each other, and the phase-shift rules can provide poor-quality results. That is why it is important to introduce a phase shift rule suitable for different Hamiltonians, solving the problem even in the degenerate case.
\par As a resource measure the number of distinct circuits that need to be evaluated to obtain all terms of a shift rule is considered. Hence, an open question is how to select the shifting parameters. Some attempts to study the different shifts are done in \cite{PhysRevA.104.052417} for a standard parameter-shift rules considering symmetric and distinct shifts.An experimental demonstration of practical on-chip PQC training with PSR is provided in \cite{10.1145/3489517.3530495}.In \cite{vidal2018calculus} the parameter shift rule is derived for the case of integer equidistant eigenvalues. However, to our best knowledge, no study is provided on the optimal selection of the phase shifts, and no analyses is done for the Hamiltonian systems with close eigenvalues.

\subsection{Contributions of this paper}
\par In this manuscript, we introduce the parameter shift rule method with shift selection to derive any order derivative and its linear combinations. Writing the unitary evolution $e^{iHt}$ as a sum of finite powers of the Hamiltonian $H$ (see Appendix~\ref{app_0}), we reduce the problem to solving the operator equation of the type:
\begin{eqnarray}\label{72_1}
    E_m(\vec{\phi})b(\vec{\phi})=\mu_m.
\end{eqnarray}
Here $E_m(\vec{\phi})$ is a $m\times m$ matrix and $\mu_m$ is a $m$ size vector, both dependent on the differences between every pair of eigenvalues $\{\lambda_i\}_{i=1}^n$ of the Hamiltonian.
To optimize the shift, one needs to solve \eqref{72_1} depending on the differences between the eigenvalue couples.
\par
\begin{figure} [t]
    \begin{center}
        \includegraphics[width=8.5cm]{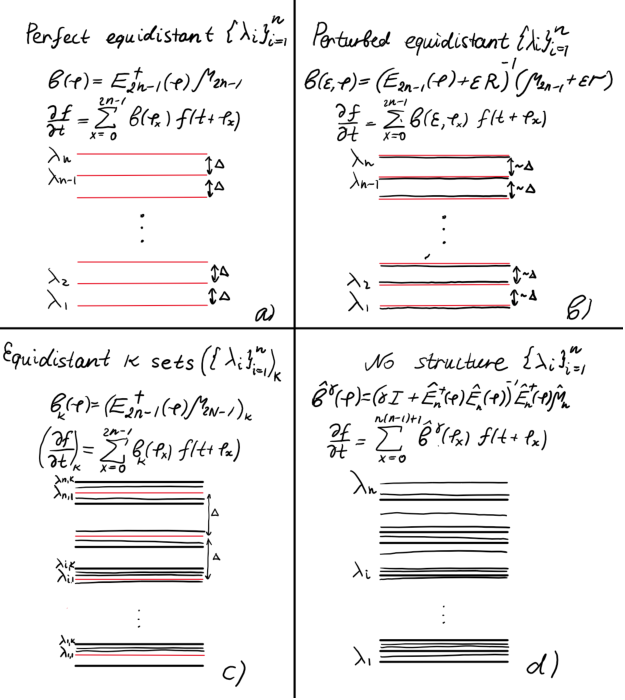}
        \end{center} 
         \caption{Different cases of the eigenvalue structures a) the equidistant eigenvalues; b) perturbed equidistant eigenvalues; c) sets of equidistant eigenvalues; d) no structure eigenvalues \label{fig.1} }
    \end{figure}

 It is known that the problem of searching for the solution of the operator equation \eqref{72_1} is called
correct by Hadamard (well-posed) if the solution exists, is unique, and is stable. If the solution does not satisfy at least one of these three conditions, it is ill-posed \cite{tikhonov1977solutions}.
In general, finding the optimal phase shifts can be an ill-posed problem by Hadamard due to the fact that some eigenvalues may be close to each other, which will give similar differences between the different couples of eigenvalues.
In the case of the well-posed problem, we provide the set of $b(\vec{\phi})$ giving the best estimate of $f'(t)$ and any of its higher derivatives $f^{(p)}(t)$ and its linear combinations. In the complicated case of the ill-posed problem, we start from the ideal case of perfectly equidistant eigenvalues (see Fig.~\ref{fig.1} a)). In this case, the matrix $E_m(\vec{\phi})$ becomes singular since some of the differences between the eigenvalues of the Hamiltonian will coincide.
We show that one can reduce the dimension of the system of linear equations to make it  well-posed solvable problem. The exact solution to the problem and the best set of phase shifts  are provided. It is interesting to mention that this solution was intuitively introduced in \cite{Wierichs2022generalparameter}. However, we prove that this is the only possible solution for such a problem, hence being the optimal one.
If the eigenvalues are not perfectly equidistant but slightly perturbed (see Fig.~\ref{fig.1} b) from the equidistant positions, one can still use the provided solution.
 
This case relates to a realistic situation since the eigenvalues are not estimated perfectly and their values can be corrupted by the estimation errors and measurement noises. The distance of the obtained solution from the optimal one is provided and is dictated by the rate of perturbation.
After we consider the case of equidistant sets of eigenvalues (see Fig.~\ref{fig.1} c)). The physical scenario is related to the previous case, corresponding to the case of different sets of experiments to estimate the eigenvalues. Unfortunately, if the system is far from the equidistant case, we can't use the latter results. The last case we consider is the most general one, where no structure in the position of the eigenvalues is detected (see Fig.~\ref{fig.1} d)).
We show how to solve such an ill-posed problem using the regularization method~\cite{tikhonov1977solutions}.
By introducing the regularization parameter that provides the possibility to find an approximate solution of \eqref{72_1} tending to the true one, we give a recipe to find the best coefficients and phase shifts numerically.
Hence, our work fully covers all the cases of Hamiltonians, providing the optimal solution for the equidistant eigenvalue case and giving a tool to find ones for non-equidistant eigenvalues.

\subsection{Organization of the paper}
\par The paper is organized as follows:
In Sec.~\ref{sec_11} we briefly recall the notion of the known parameter-shift rules.
In Sec.~\ref{sec_3} we discuss the general parameter shift rule and deduce the optimal coefficient for a well-posed problem.
In Sec.~\ref{sec_4} we study the ill-posed problem. The cases of equidistant eigenvalues of the Hamiltonian, equidistant eigenvalues except for one, slightly perturbed equidistant eigenvalues, and highly non-equidistant eigenvalues forming equidistant sets are studied in detail.
In Sec.~\ref{sec_5} the phase-shift solution for an ill-posed by Hadamard problem for no-structure eigenvalues is proposed.
We end the main text in Sec.\ref{sec_7} with a discussion. Finally, in the appendix, we
summarize some technical derivations.

\section{Overview of Known Parameter-Shift Rules}\label{sec_11}
\par Let $\ket{\psi}$ denote the quantum state in the Hilbert space. Consider the unitary operator $U(t)=e^{iHt}$, defined by a Hamiltonian $H$ and a parameter $t$.
The eigenvalues of $U(t)$ are given by $\{\exp(i\lambda_j t)\}_{j\in 1}^n$ with real-valued
$\{\lambda_j\}_{j\in 1}^n$ and have
sorted the $j$ to be non-decreasing.
We are interested in the mean value of a measurable observable $C$ defined as follows:
\begin{equation}\label{1249}
 f(t) \equiv \bra{\psi}U(t)^{\dagger} C U(t)\ket{\psi}.
\end{equation}
The expectation value $f(t)$ can be written as a finite-term Fourier series
\begin{eqnarray}
\label{1506}f(t)=a_0+\sum\limits_{l=1}^{m} a_l\cos{(\Omega_l t)}+b_l\sin{(\Omega_l t)},
\end{eqnarray}
where we have $m$ unique positive differences $\{\Omega_l\}_{l\in m}=\{|\lambda_j-{\lambda}_k|, j,k\in[1,n],{\lambda}_j>\lambda_k\}$.
\par For functions $f(t)$ with a single frequency
$\Omega_1=\Omega$ (i.e., $U$ has two eigenvalues), the derivative
can be computed via the parameter-shift rule \cite{li2017hybrid,mitarai2018quantum,schuld2019evaluating}:
\begin{eqnarray}
 \frac{\partial f(t)}{\partial t}\Big|_{t=0}=\frac{\Omega}{2\sin{(\Omega \phi_1)}}(f(\phi_1)-f(-\phi_1)),\quad \phi_1\in (0,\pi).
\end{eqnarray}
In \cite{anselmetti2021local} the latter rule is generalized to gates with eigenvalues
$\{-1, 0, 1\}$, which leads to $m = 2$ frequencies:
\begin{eqnarray}\label{1228}
\frac{\partial f(t)}{\partial t}\Big|_{t=0}=y_1(f(\phi_1)-f(-\phi_1))-y_2(f(\phi_2)-f(-\phi_2)),\quad \phi_1,\phi_2\in (0,\pi),
\end{eqnarray}
and $y_{1,2}$ are the corresponding coefficients. In \cite{anselmetti2021local} $\phi_{1,2}=\pi/2\mp \pi/4$ and $y_{1,2}=(\sqrt{2}\pm 1)/2\sqrt{2}$ is studied.
On the other hand, in \cite{kottmann2021feasible} for the same eigenvalues, the following rule is introduced:
\begin{eqnarray}
 \frac{\partial f(t)}{\partial t}\Big|_{t=0}=\frac{1}{4}(f_+^+-f_-^++f_+^--f_-^-),
\end{eqnarray}
where $f_{\pm}^{\alpha}$ is the measured energy when replacing the
gate $U(t)$ in question by $U(t\pm \pi/2) \exp(\mp \alpha i \pi/4 \Pi_0)$, where $\Pi_0$
is the projector onto the zero-eigenspace of
the generator of $U$.
\par For the perturbed quantum evolution $U_F(t)=\exp{(i(t H+F)}$ the stochastic parameter-shift rule is introduced in \cite{banchi2021measuring}
\begin{eqnarray}\label{1229}
\frac{\partial f(t)}{\partial t}\Big|_{t=\phi_0}=\frac{\Omega}{2\sin{(\Omega \phi_1)}}\int\limits_{0}^{1}(f_+(r)-f_-(r)) dr,
\end{eqnarray}
where $f_{\pm}(t)$ is the energy measured in the state prepared
by a modified circuit that splits $U_F(r_0)$ into
$U_F(r \phi_0)$ and $U_F((1-r)\phi_0)$, and interleaves these two
gates with $U_{F=0}(\pm \phi_0)$.
These results were further developed to introduce the Nyquist shift rule in \cite{Theis2023propershiftrules}.
A parameter-shift rule for higher-order derivatives
based on repeatedly applying the original rule,
been proposed in \cite{PhysRevA.103.012405}.
\par In \cite{Wierichs2022generalparameter} the so-called general parameter shift rules are defined for the case of evenly spaced phase shifts $\phi_j=(2j-1)\pi/2n$ ($\phi_j=j\pi/n$), $j\in \overline{1,n}$ to reconstruct odd (even) functions:
\begin{eqnarray}
 f'(0)&=&\sum\limits_{j=1}^{2n} f\left(\frac{(2j-1)\pi}{2n}\right)\frac{(-1)^{j-1}}{4n \sin^2{(\frac{(2j-1)\pi}{4n})}},\\\nonumber
f''(0)&=&-f(0)\frac{2n^2+1}{6}+\sum\limits_{j=1}^{2n-1} f\left(\frac{j\pi}{n}\right)\frac{(-1)^{j-1}}{2 \sin^2{(\frac{j\pi}{2n})}},\\\nonumber
\end{eqnarray}
The latter result coincides with \eqref{1228} and \eqref{1229} in parameter-shift rules for $n = 1$ and $n = 2$, respectively.
\par The selection of phase shifts  depends on various factors, including the specific problem being solved, the available resources, and the desired accuracy. Phase shifts may be chosen based on mathematical considerations or analytical insights into the problem structure. In other cases, numerical methods or optimization techniques can be employed to find optimal phase shift values that minimize errors or maximize efficiency. Further, we introduce the optimal parameter shift selection method suitable for any structure of the Hamiltonian system.

\section{Parameter Shift Rule for a Well-Posed Problem }\label{sec_3} 
Any function from $H$ can be written as a finite sum of $H$ powers (see Appendix~\ref{app_0}), namely 
\begin{eqnarray}
e^{\mathbbm{i} H t} = \sum_{k=0}^{n-1} a_k(t) H^k,\end{eqnarray} 
hold. Here the coefficients are $\ket{a(t)} = \Lambda^{-1} \ket{e(t)}$, where $\Lambda$ is the $n
\times n$ Vandermonde matrix containing the $\vec \lambda$, and $\langle{k|e(t)}\rangle = e^{\mathbbm{i}\hat{\mu}_k t}$, holds.
Then, we can rewrite \eqref{1249} as follows:
\begin{equation}\label{1357}
    f(t) = \mathrm{Tr}[\ket{e(t)}\bra{e(t)} (\Lambda^{-1})^\dagger \tilde{C} \Lambda^{-1}],
\end{equation}
where $(\tilde{C})_{k,l} \equiv  \bra{\psi}H^kC H^l\ket{\psi}$. 
\par The parameter-shift rules relate derivatives of a quantum
function to evaluations of the function itself at different points.
Using \eqref{1357}, we can rewrite the PSR \eqref{0}  as follows:
\begin{equation}
    \ket{e'(t)}\bra{e(t)} - \ket{e(t)}\bra{e'(t)} = \sum_{x=1}^{m} b_x(\vec{\phi}) \ket{e(t+\phi_x)}\bra{e(t+\phi_x)}.
\end{equation}
The latter is equivalent to solving the following system of equations
\begin{equation}
    \mathbbm{i}(\lambda_k - \lambda_l)e^{\mathbbm{i}(\lambda_k - \lambda_l)t} = \sum_{x=1}^{m} b_x (\vec{\phi})e^{\mathbbm{i}(\lambda_k - \lambda_l)t} e^{\mathbbm{i}(\lambda_k - \lambda_l)\phi_x},\quad \forall k,l \in \overline{1,n},
\end{equation}
where we used the notation $\langle (k,l)|{e}(t)\rangle = e^{\mathbbm{i}({\lambda}_k-\lambda_l) t}$.
Since the latter equation must be satisfied for every $t$, the compatibility equation reads as 
\begin{equation}\label{948}
       \sum_{x=1}^{m} b_x(\vec{\phi}) e^{\mathbbm{i}(\lambda_k - \lambda_l)\phi_x} =  \mathbbm{i}(\lambda_k - \lambda_l),\quad  \ \forall k, l \in \overline{1,n}.
\end{equation}
This system is highly nonlinear in the $\phi$ variables but it is nevertheless linear in the $b$'s.
\par Let $U$ and $V$ be Hermitian metric spaces  with metrics $\rho_U$ and $\rho_V$. The continues  one-to-one operator $E$  from $U$ to $V$  corresponds to  the $m\times m$ matrix $E_m$ such that its elements are $E_{(k,l),x} = e^{\mathbbm{i} \mu_{(k,l)}\phi_x}$, $ k, l \in \overline{1,n}$. Here we introduce the distance between two eigenvalues as $\mu_{(k,l)}\equiv \lambda_{k}-\lambda_{l}$, $k,l=\overline{1,n}$. 
The function $\mu$ corresponds to the $m\times 1$ vector with elements $ \mathbbm{i}(\lambda_k - \lambda_l)$. Hence, \eqref{948} in the  operator form is
\begin{eqnarray}\label{72}
    E(\vec{\phi}) b(\vec{\phi})=\mu, \quad b\in U, \quad \mu\in V.
\end{eqnarray}
\par In this section, we assume that the phases are selected in such a way that the problem is well-posed by Hadamard. The constraints on them we observe further. Then we can determine the vector of solutions of \eqref{72} as follows:
\begin{equation}
\label{parametricshiftrulegeneral}
    b(\vec{\phi}) =  E^{-1}(\vec{\phi}) {\mu}.
\end{equation}

Using Cramer's rule, the closed-form expression for every element $b_x(\vec{\phi})$ can be written as
\begin{eqnarray}\label{1230}
   b_x(\vec{\phi})=\det E(\vec{\phi}/\phi_x)\cdot (\det E(\vec{\phi}))^{-1},
\end{eqnarray}
where 
\begin{eqnarray*}
 &&   E(\vec{\phi}/\phi_x)=
    \begin{bmatrix}
1 & 1 &\dots& 0 & \dots& 1  \\
e^{i\mu_{(1,2)}\phi_1}&e^{i\mu_{(1,2)}\phi_2}&\dots & i\mu_{(1,2)}& \dots& e^{i\mu_{(1,2)}\phi_m} \\
e^{-i\mu_{(1,2)}\phi_1} &e^{-i\mu_{(1,2)}\phi_2} &\dots& -i\mu_{(1,2)}& \dots& e^{-i\mu_{(1,2)}\phi_m}  \\
\vdots &\vdots&\vdots& \vdots&  \ddots& \vdots\\
e^{i\mu_{(n-1,n)}\phi_1}&e^{i\mu_{(n-1,n)}\phi_2}&\dots & i\mu_{(n-1,n)}& \dots& e^{i\mu_{(n-1,n)}\phi_m}\\
e^{-i\mu_{(n-1,n)}\phi_1}&e^{-i\mu_{(n-1,n)}\phi_2}&\dots & -i\mu_{(n-1,n)}& \dots& e^{-i\mu_{(n-1,n)}\phi_m}
\end{bmatrix}
\end{eqnarray*}
is the matrix $E(\vec{\phi})$ with the $x$ column substituted by the $\mu$ vector and not depending on $\phi_x$.
The determinants are not equal to zero if the matrix does not contain equal columns or rows or all zero columns or rows. To this end,
we get that $\phi_i\neq \pm \phi_j+2\pi c$, $c\in Z$ and all distances $\mu_{(k,l)}$ must be different. That means that the equidistant $n$ eigenvalue problem is ill-posed by Hadamard. The solution to this problem will be provided in the next section.
\par Using Jacobi's formula, we get the alternative expression
\begin{equation}
    b_x(\vec{\phi}) = \frac{\partial{\det E(\vec{\phi})}}{\partial \phi_x}|_{\phi_x = 0}\cdot (\det E(\vec{\phi}))^{-1}.
\end{equation}
The solution is exact when $m=n(n-1) +1$, holds. This number is obtained by counting every $\mu_{(k,l)}$, $k \neq l$. The $k=l$ case always yields the same equation $\sum_x b_x (\vec{\phi})= 0$.
\par We can write $b_x(\vec{\phi})$ as follows:
\begin{equation}
    b_x(\vec{\phi}) = \frac{\left|
    \begin{array}{ccccc}
    |&&|&&|\\
    \vec{v}(\phi_0)&\cdots&\frac{\partial \vec{v}(\phi_x)}{\partial \phi_x} |_{\phi_x = 0}&\cdots &\vec{v}(\phi_{m})\\
    |&&|&&|
    \end{array}
    \right|}{
    \left|
    \begin{array}{ccccc}
    |&&|&&|\\
    \vec{v}(\phi_0)&\cdots& \vec{v}(\phi_x)&\cdots &\vec{v}(\phi_{m})\\
    |&&|&&|
    \end{array}
    \right|
    },\quad x\in\overline{1,m},
\end{equation}
where 
the vectors forming the matrix $E(\vec{\phi})$ are denoted as 
\begin{equation}\label{1146}
    \vec{v}(\phi_i)=
    \begin{pmatrix}
        1 & \exp(\mathbbm{i} \mu_{(12)} \phi_i) & \exp(-\mathbbm{i} \mu_{(12)} \phi_i) & \cdots & \exp(\mathbbm{i} \mu_{(n-1,n)} \phi_i) & \exp(-\mathbbm{i} \mu_{(n-1,n)}\phi_i)
    \end{pmatrix},
\end{equation}
This automatically yields a PSR for the derivatives of arbitrary order without increasing the number of function evaluations. The following Theorem holds:
\begin{thm}\label{th1} Let $n$ be the number of distinct, not equidistant, eigenvalues of $H$. Let $\vec{\phi} \in \mathbbm{R}^m$ with $m = n(n-1) +1$ and $\phi_i\neq \pm \phi_j+2\pi c$, $c\in Z$, $\forall i,j\in \overline{1,m}$. Then, the following parameter shift rule
\begin{equation}
    \frac{\partial^p f(t)}{\partial t^p} = \sum_{x=0}^{m-1} b_x^{(p)}(\vec{\phi}) f(t+ \phi_x),\quad p\geq 1,
\end{equation}
holds, if, and only if, the vector $b_x(\vec{\phi}) $ satisfies
\begin{equation}\label{1444}
        b_x^{(p)}(\vec{\phi}) = \frac{\left|
    \begin{array}{ccccc}
    |&&|&&|\\
    \vec{v}(\phi_0)&\cdots&\frac{\partial^p\vec{v}(\phi_x)}{\partial \phi_x^p} |_{\phi_x = 0}&\cdots &\vec{v}(\phi_{m})\\
    |&&|&&|
    \end{array}
    \right|}{
    \left|
    \begin{array}{ccccc}
    |&&|&&|\\
    \vec{v}(\phi_0)&\cdots& \vec{v}(\phi_x)&\cdots &\vec{v}(\phi_{m})\\
    |&&|&&|
    \end{array}
    \right|
    }.
\end{equation}
\end{thm}
\textit{Proof:} Left as an exercise.

The latter statement can  be generalized. In particular, any linear combination of high-order derivatives can be expressed similarly as follows:

\begin{gather}
    \sum_i a_i f^{(i)}(t)=\sum_{x=0}^{m-1} \tilde b_x(\vec{\phi}) f(t+\phi_x),\quad  \text{where}\\\nonumber
    \tilde b_x(\vec \phi)
    =
    \frac{\left|
    \begin{array}{ccccc}
    |&&|&&|\\
    \vec{v}(\phi_0)&\cdots&\sum_i a_i \frac{\partial^i\vec{v}(\phi_x)|_{\phi_x = 0}}{\partial \phi_x^i} &\cdots &\vec{v}(\phi_{m})\\
    |&&|&&|
    \end{array}
    \right|}{
    \left|
    \begin{array}{ccccc}
    |&&|&&|\\
    \vec{v}(\phi_0)&\cdots& \vec{v}(\phi_x)&\cdots &\vec{v}(\phi_{m})\\
    |&&|&&|
    \end{array}
    \right|
    },
\end{gather}
holds.
Even though it is not generalizable for any other algebraic expression $F=F(f,f',f''\dots)$ involving non-linear terms, we can always re-write it as products of functions we can compute. 

\par The Theorem~\ref{th1}  works for any shift vector $\vec{\phi}$ such that the problem \eqref{parametricshiftrulegeneral} is well posed. 
The variance of the estimate of the derivative $\widehat{\frac{\partial f(t,\vec{\phi})}{\partial t}}$ is
\begin{equation}\label{1219}
\sigma^2\left(\widehat{\frac{\partial f(t,\vec{\phi})}{\partial t}}\right) = \sum_{x=1}^{m} b_x^2(\vec{\phi}) \sigma^2 (\hat{f}(t+ \phi_x)).
\end{equation}
The Chebyshev's inequality can be written
\begin{eqnarray}
 \mathbb{P}\left(\Bigg|{\frac{\partial f(t)}{\partial t}}-\widehat{\frac{\partial f(t,\vec{\phi})}{\partial t}}\Bigg|\geqslant \nu\right) \leqslant \frac{1}{ \nu^2}\sum_{x=1}^{m} b_x^2(\vec{\phi})\sigma^2 (\hat{f}(t+ \phi_x)),\end{eqnarray}
where $\nu>0$ is a real number. For the probability $\eta\in(0,1)$, we get
\begin{eqnarray}
    \nu=\left(\frac{1}{ \eta}\sum_{x=1}^{m} b_x^2(\vec{\phi})\sigma^2 (\hat{f}(t+ \phi_x))\right)^{\frac{1}{2}}.
\end{eqnarray}
Then with probability $\eta$ the confidence interval for the estimate of the derivative is
\begin{eqnarray}
 {\frac{\partial f(t)}{\partial t}}-\nu\leq    \frac{\partial \widehat{f(t,\vec{\phi})}}{\partial t}\leq {\frac{\partial f(t)}{\partial t}}+\nu.
\end{eqnarray}
We take the derivative of the variance by $\phi_y\in\vec{\phi}$, $y=\overline{1,m}$.
Equating it to zero, we get the condition 
\begin{eqnarray}
     2\sum_{x=1}^{m} b_x(\vec{\phi})\frac{\partial b_x(\vec{\phi})}{\partial \phi_y} \sigma^2 (\hat{f}(t+ \phi_x))
   =-b_y^2(\vec{\phi})\frac{\partial \sigma^2 (\hat{f}(t+ \phi_y))}{\partial \phi_y}.
\end{eqnarray}
We can assume that  the variances $\sigma^2 (\hat{f}(t+ \phi_y))$ are not dependent on the phase and are equal. Then we get
\begin{eqnarray}\label{1255}
     \sum_{x=1}^{m} b_x(\vec{\phi})\frac{\partial b_x(\vec{\phi})}{\partial \phi_y}=\mathbf{0},\quad \forall y\in \overline{1,m}.
\end{eqnarray}
We can rewrite \eqref{1255} as (see Appendix~\ref{ap_3})
\begin{eqnarray} \label{1508}\sum\limits_{x=1}^{m}\det(E(\vec{\phi}/\phi_x))
\det( E_y(\vec{\phi}/\phi_x))
   =(\det E(\vec{\phi}))^{-1}\det( E_y(\vec{\phi}))\sum\limits_{x=1}^{m}\left(\det(E(\vec{\phi}/\phi_x))\right)^2
.
\end{eqnarray}
Solving the latter system of equations with respect to all $\phi_y$, $y\in\overline{1,m}$, one can find the optimal $\vec{\phi}$ minimizing the variance \eqref{1219}.

\section{Optimal Phase Shift Parameters Selection for Ill-Posed Problem}\label{sec_4}
\par In this section, we assume that the problem \eqref{72} is ill-posed. In this case, the solution \eqref{parametricshiftrulegeneral} is unstable.
\par Let us first look for a set of phase shifts such that the vectors forming the matrix $E_m$
\begin{equation}\label{1146_10}
    \vec{v}(\phi_i)=
    \begin{pmatrix}
        1 & \exp(\mathbbm{i} \mu_{(12)} \phi_i) & \exp(-\mathbbm{i} \mu_{(12)} \phi_i) & \cdots & \exp(\mathbbm{i} \mu_{(n-1,n)} \phi_i) & \exp(-\mathbbm{i} \mu_{(n-1,n)}\phi_i)
    \end{pmatrix},
\end{equation}
would be orthogonal to each other. Here we use that $\mu_{(k,l)}=-\mu_{(l,k)}$ holds, so we use the notation $\mu_{(t,p)}$, $t<p$, $\forall t,p\in \overline{1,n}$. In this case, the inversion of $E_m$ is equal to its hermitian conjugation.
\par We impose the orthogonality condition on the columns of $E_m$, namely
$\vec{v}^{\star}(\phi_j)\vec{v}(\phi_i)=0$. Then we get
\begin{equation}\label{1301}
    1+2\sum_{
    t,p=1,t<p}^{n}\cos\big(\mu_{(t,p)} (\phi_i-\phi_j)\big)=0 , \quad \forall \phi_i,\phi_j\in\vec{\phi}.
\end{equation}
Let us denote the difference between phase shifts as $\Phi_{ij}\equiv \phi_i-\phi_j$ and rewrite the latter condition as follows
\begin{eqnarray}
    \label{1301_1}
    &&1+2\sum_{
    p=2}^{n}\cos\big(\mu_{(1,p)}\Phi_{ij}\big)+2\sum_{
    p=3}^{n}\cos\big(\mu_{(2,p)}\Phi_{ij}\big)+\dots \\\nonumber
    &\dots&+2\sum_{
    p=m-1}^{n}\cos\big(\mu_{(n-2,p)} \Phi_{ij}\big)+2\cos\big(\mu_{(n-1,n)} \Phi_{ij}\big)=0.
\end{eqnarray}
To solve the latter equation, we need to make some assumptions on the eigenvalue distances $\mu_{t,p}$. Below, we first discuss the equidistant Hamiltonian eigenvalues case, moving on to the perturbed case in the following subsection.
\subsubsection{Equidistant Eigenvalues}
\par Let us assume that all eigenvalues $\{\lambda_i\}_{i=1}^n$ are equidistant (see Fig.\ref{fig.2} a) and denote the distance between two neighboring eigenvalues as $\Delta$.
One can see that $\mu_{(1,2)}=1\Delta$, $\mu_{(1,3)}=2\Delta$ and $\mu_{(1,n)}=(n-1)\Delta$. Similarly, $\mu_{(2,4)}=2\Delta$ and $\mu_{(2,n)}=(n-2)\Delta$, hold. So, we can conclude that $\mu_{(t,p)}=(p-t)\Delta$, $t<p$. One can see that in this case some rows of the matrix $E_m$ will coincide and it will become singular. The problem \eqref{72} is ill-posed.
\par Since we are interested in the inversion of the matrix $E_m$, we exclude all the similar rows, reducing the matrix $E_m$ to the matrix of a smaller size $E_{2n-1}$ which is non-singular:
\begin{eqnarray}\label{1458}
  \!\!\!  \!\!\!E_{2n-1}(\vec{\phi})=
\begin{bmatrix}
1 & 1 &  \dots& 1  \\
e^{\mathbbm{i}1\Delta\phi_1} &e^{\mathbbm{i}1\Delta\phi_2} &  \dots& e^{\mathbbm{i}1\Delta\phi_{2n-1}} \\
e^{-\mathbbm{i}1\Delta\phi_1} &e^{-\mathbbm{i}1\Delta\phi_2} & \cdot& e^{-\mathbbm{i}1\Delta\phi_{2n-1}}  \\
\vdots &\vdots&   \ddots& \vdots\\
e^{-\mathbbm{i}(n-1)\Delta\phi_1} &e^{-\mathbbm{i}(n-1)\Delta\phi_2} &  \dots& e^{-\mathbbm{i}(n-1)\Delta\phi_{2n-1}}
\end{bmatrix},\!\!\!\quad \vec{\mu}_{2n-1}=\mathbbm{i}\Delta\begin{bmatrix}
0\\
1 \\
-1\\
\vdots \\
-(n-1)
\end{bmatrix}.
\end{eqnarray}
Here $\vec{\mu}_{2n-1}$ is a vector of all unique distances $\mu_{(1,i)}$, $i=\overline{1,n}$.
In this case the condition \eqref{1301_1} can be reduced to the following one
\begin{eqnarray}\label{1540}
    1+2\sum_{
    k=1}^{n-1}\cos\big(k\Delta\Phi_{ij}\big) =0.
\end{eqnarray}
The Dirichlet kernel is  defined as follows
\begin{eqnarray}
     D_n(x) = 1 + 2\sum_{k=1}^{n} \cos{(kx)}
= \frac{\sin\left(\left(n + \tfrac12 \right)x\right)}{\sin{(\tfrac12 x)}},\end{eqnarray}
where its zeros are at the points $x_t=\frac{2\pi t}{2n+1}$, $t\in \mathbbm{Z}$.
Hence, the condition \eqref{1540} can be rewritten as
\begin{eqnarray}
   D_{n-1}(\Delta\Phi_{ij}) =0,
\end{eqnarray}
and the solution is given by
\begin{eqnarray}
   \Delta\Phi_{ij}=\frac{2\pi t_{ij}}{2n-1},\quad t_{ij}\in \mathbbm{Z}.
\end{eqnarray}
Finally, we have a system of equations
\begin{eqnarray}\label{1456}
   \phi_{i}-\phi_{j}=\frac{2\pi }{(2n-1)\Delta} t_{ij},\quad t_{ij}\in Z,\quad \forall i,j=[1,{2n-1}],\quad  i<j.
\end{eqnarray}
To solve the latter system of equations we first consider the case of equidistant phase-shifts $\phi_j$, $\forall j$. From \eqref{1456} we conclude:
\begin{eqnarray}\label{1456_2}
   \phi_{j}-\phi_{j+1}=\frac{2\pi }{(2n-1)\Delta},\quad \forall j=[1,2n-2],\quad  t_{j,j+1}=1.
\end{eqnarray}
It is straightforward to verify, that the solution of the latter system in the equidistant phases case is given by
\begin{eqnarray}\label{1234}
    \phi_j=-\frac{2\pi j}{(2n-1)\Delta},\quad \forall j=[1,{2n-1}].
\end{eqnarray}
One can see that if $j=2n-1$ holds, then $\phi_{2n-1}=-2\pi/\Delta$,  and $\exp{j\Delta \phi_{2n-1}}=1$. 
Then the matrix \eqref{1458} reduces to
\begin{eqnarray}\label{1459_10}
    E_{2n-1}=
\begin{bmatrix}
1 & 1 & 1 & \dots& 1 \\
e^{-i\tau} &e^{-2i\tau} & e^{-3i\tau}& \dots& 1 \\
e^{i\tau} &e^{2i\tau} & e^{3i\tau}& \dots& 1 \\
e^{-2i\tau} &e^{-4i\tau} & e^{-6i\tau}& \dots& 1 \\
e^{2i\tau} &e^{4i\tau} & e^{6i\tau}& \dots& 1 \\
\vdots &\vdots& \vdots&  \ddots& \vdots\\
e^{-(n-1)i\tau} &e^{-2(n-1)i\tau} & e^{-3(n-1)i\tau}& \dots& 1\\
e^{(n-1)i\tau} &e^{2(n-1)i\tau} & e^{3(n-1)i\tau}& \dots& 1\\
\end{bmatrix},
\end{eqnarray}
where we used the notation $\tau=\frac{2\pi}{(2n-1)}$. Let us normalise the latter matrix, introducing 
\begin{eqnarray}
    \tilde{E}_{2n-1}\equiv  E_{2n-1}/\sqrt{2n-1}.\end{eqnarray} This matrix is unitary since its rows and columns are orthonormal.
    Then its inverse matrix is $\tilde{E}_{2n-1}^{\dagger}$
    and the solution of our problem \eqref{parametricshiftrulegeneral}  is the following
    \begin{eqnarray}
     {b}_{2n-1}(\phi)= \tilde{E}_{2n-1}^{\dagger}\vec{\tilde{\mu}}_{2n-1},\quad \vec{\tilde{\mu}}_{2n-1}\equiv \frac{1}{\sqrt{2n-1} } \vec{\mu}_{2n-1}.
    \end{eqnarray}
Since we know the form of the matrix \eqref{1459} explicitly, we can write the solution:
   \begin{eqnarray}\label{1615}
     {b}_{2n-1}(\tau,\Delta)=-\frac{2 \Delta}{{2n-1}}\begin{bmatrix}\sum\limits_{j=0}^{n-1}2^j\sin{((j+1)\tau)}\\
\sum\limits_{j=0}^{n-1}2^j\sin{((j+1)2\tau)}\\
\sum\limits_{j=0}^{n-1}2^j\sin{((j+1)3\tau)}\\
\vdots \\
\sum\limits_{j=0}^{n-1}2^j\sin{((j+1)(2n-2)\tau)}\\0
\end{bmatrix}.
    \end{eqnarray}  

\par Hence, we found the explicit solution to our problem in the case of equidistant eigenvalues and phase shifts. However, is it possible to solve the system \eqref{1456} without imposing the latter constraint? The general solution is provided in Appendix~\ref{ap_1}. However, due to the periodicity of the complex exponent, using this solution, the matrix $E_{2n-1}$ becomes singular in all cases except for equidistant phases one. 
\par We can conclude that in the case of equidistant eigenvalues, the amount of unique distances between them reduces to $2n-1$ and only the equidistant phase shifts given by \eqref{1234} guarantee non-singularity of $E_{2n-1}$. In this case, the system \eqref{72} has a unique solution \eqref{1615}. To find the function derivative, one needs $2n-2$ phase shifts, where $n$ is the number of eigenvalues.
\subsubsection{Equidistant Eigenvalues Except of One}
\par Let us assume that all eigenvalues $\{\lambda_i\}_{i=2}^{n}$ are equidistant with the distance between every neighboring one denoted by $\Delta$, namely $\mu_{(2,3)}=1\Delta$, $\mu_{(2,4)}=2\Delta$ and $\mu_{(2,n)}=(n-2)\Delta$. The first one is distant from all the others, where $\mu_{(1,2)}=\Delta_1$, $\mu_{(1,3)}=\Delta_1+\Delta$,  $\mu_{(1,4)}=\Delta_1+2\Delta$ and $\mu_{(1,n)}=\Delta_1+(n-2)\Delta$. Then \eqref{1301_1} is reducing to
\begin{eqnarray}
    \label{1301_1_1}
    &&1+2\cos\big(\Delta_1\Phi_{ij}\big)+2\sum_{
    k=1}^{n-2}\left[\cos\big(k\Delta\Phi_{ij}\big)+\cos\big((\Delta_1+k\Delta)\Phi_{ij}\big)\right]=0.
\end{eqnarray}
It can be rewritten as
\begin{eqnarray}
    \label{1301_1_2}
    \frac{1}{2}+\cos\big(\Delta_1\Phi_{ij}\big)+\sum_{
    k=1}^{n-2}\left[\cos\big(k\Delta\Phi_{ij}\big)(1+
    \cos\big(\Delta_1\Phi_{ij}\big))- \sin\big(\Delta_1\Phi_{ij}\big)
    \sin\big(k\Delta\Phi_{ij}\big)\right]=0.
\end{eqnarray}
According to the definition of the Dirichlet and the conjugate Dirichlet  kernels the latter expression can be rewritten as
\begin{eqnarray}
    \label{1301_1_5}
    &&\left(1+\cos\big(\Delta_1\Phi_{ij}\big)\right)D_{n-2}\left(\Delta\Phi_{ij}\right)+\frac{1}{2}\cos\big(\Delta_1\Phi_{ij}\big)=\sin\left(\Delta_1\Phi_{ij}\right)\tilde{D}_{n-2}\left(\Delta\Phi_{ij}\right).
\end{eqnarray}
The general solution to the latter expression can be found, having the form $
\Delta_1 \Phi_{ij}= f(n,\Delta \Phi_{ij})$, where $f(\cdot)$ is a combination of trigonometric functions.
One can see that in this case $\Delta_1$ is different for every $\Phi_{ij}$, however, the distance must be the same $\forall i,j$ pairs of phases.
\par For example, one of the solutions to the latter equation is
\begin{eqnarray}
   &&\Phi_{ij}=\frac{2\pi t_{ij}}{(2n-3)  \Delta},\quad t_{ij}\in \mathbbm{Z},\\\nonumber
   &&\Delta_1= \frac{(2n-3)}{2 t_{ij}}(\cot ^{-1}\left(\frac{1}{2} \left(\cos \left(\frac{2 \pi  t_{ij}}{3-2 n}\right)+(-1)^{t_{ij}+1}\right) \csc \left(\frac{\pi  t_{ij}}{2 n-3}\right)\right)+\pi  c)\Delta, \quad c\in \mathbbm{Z},
\end{eqnarray}
meaning equidistant phases and the distance of the out eigenvalue to scale with $n>2$.
However, it is not possible to find all the phase shifts since $\Delta_1$ is dependent on $t_{ij}$ which is different for every $\Phi_{ij}$. This is contradictory to the fact that $\Delta_1$ must be constant.
\par We can conclude that in the case of all equidistant eigenvalues except one, the orthogonality condition on the vectors forming the matrix $E_m$ is not fulfilled. That means that the orthogonality property is a specific feature of equidistant eigenvalue systems.

\subsubsection{Slightly Perturbed Equidistant Eigenvalues}\label{sec_1}
\par An equidistant eigenvalue case is a theoretical assumption that is not the case in any of the realistic scenarios. However, the eigenvalues can be close to the ideal equidistant positions. This case can be treated using the perturbation theory~\cite{golub2013matrix}.
\par Let us perturb the equidistant system
$\tilde{E}_{2n-1}(\vec{\phi})b(\vec{\phi})= \vec{\tilde{\mu}}_{2n-1}$,  namely
\begin{equation}\label{2}
   (\tilde{E}_{2n-1}(\vec{\phi})+\varepsilon \tilde{R}_{2n-1}(\vec{\phi}))b(\varepsilon,\vec{\phi})=\vec{\tilde{\mu}}_{2n-1}+\varepsilon \tilde{r}_{2n-1}.
   \end{equation}
   This corresponds to the case when the eigenvalues of the Hamiltonian are not ideally equidistant but  slightly shifted from equidistant positions. In Appendix~\ref{ap_2}  we deduce the perturbation matrices to be
   \begin{eqnarray}
   && \tilde{R}_{2n-1}\equiv \frac{ R_{2n-1}}{\sqrt{2n-1}} ,\quad \tilde{r}_{2n-1}\equiv \frac{\mathbbm{i}  I_{2n-1}}{\sqrt{2n-1}},\\\nonumber
&&R_{2n-1}=\frac{\mathbbm{i}\tau}{\Delta}
\begin{bmatrix}
0 & 0 & \dots& 0  \\
e^{-\mathbbm{i}\tau} &2e^{-2\mathbbm{i}\tau}  & \dots& (2n-1)  \\
-e^{\mathbbm{i}\tau} &-2e^{2\mathbbm{i}\tau}& \cdot& - (2n-1)  \\
\vdots &\vdots&   \ddots& \vdots\\
-e^{\mathbbm{i}(n-1)\tau}&-2e^{\mathbbm{i}(n-1)\tau}& \dots& - (2n-1)
\end{bmatrix}.
\end{eqnarray}
Here $\varepsilon>0$ is a perturbation parameter.
For a nonsingular matrix $E_{2n-1}$ the perturbed matrix $E_{2n-1} + \varepsilon R_{2n-1}$ is also nonsingular if the perturbation $\varepsilon R$ is sufficiently small. Further in this subsection, we omit the $\vec{\phi}$ in the brackets, $2n-1$ subscripts and the $\sim$ superscript.
\par Differentiating by $\varepsilon$ (we suppose that this derivative exists), one can derive
\begin{equation}\label{3}
   E\dot{b}(\epsilon)+Rb(\varepsilon)+\varepsilon R\dot{b}(\varepsilon)= r.
   \end{equation}
Then for $\epsilon=0$ we get the following expression
\begin{equation}\label{4}
   E\dot{b}(0)+Rb(0)= r\longrightarrow \dot{b}(0)=E^{-1}(r-Rb(0)).
   \end{equation}
   Note that $b(0)$ is the solution \eqref{1615} of the not-perturbed problem.   Using the Taylor expansion
   \begin{equation}\label{5}
  {b}(\varepsilon)= {b}(0)+\varepsilon \dot{b}(0)+o(\varepsilon),
   \end{equation}
  we can write
     \begin{eqnarray}\label{6}
  &&\frac{\|b(\varepsilon)-{b}(0)\|}{\|b(0)\|}=\epsilon \frac{\|E^{-1}(r-Rb(0))\|}{\|b(0)\|}+o(\varepsilon)\leq  \|E^{-1}\|\left(\frac{\|\varepsilon r\|}{\|b(0)\|}+\|\varepsilon R\|\right)+o(\varepsilon)\nonumber\\
  &=&   \|E^{-1}\|\|E\|\left(\frac{\|\varepsilon r\|}{\|Eb(0)\|}+\frac{\|\varepsilon R\|}{\|E\|}\right)+o(\epsilon)\leq k(E)\left(\frac{\|\varepsilon r\|}{\|{\vec{\mu}}\|}+\frac{\|\varepsilon R\|}{\|E\|}\right)+o(\varepsilon),
   \end{eqnarray}
where $k(E)\equiv \|E^{-1}\|\|E\|\geq 1$ is the condition number.
An ill-conditioned system is one with a large condition number. If the system is ill-conditioned,
then a small perturbation to the RHS can lead to large changes in the solution.
When $k(E)$ is large, this implies that $b(\varepsilon)$ can be very far from $b(0)$.
\par The distance between the solutions is given by
\begin{eqnarray}
   \|b(\varepsilon)-{b}(0)\|\approx \|E^{-1}\|\left(\|\varepsilon r\|+\|\varepsilon R\|\right)\|b(0)\|.
\end{eqnarray}
In the case of equidistant eigenvalues and phases, which we discussed in the previous section, the matrix $\tilde{E}_{2n-1}$ is unitary. The norm of the unitary matrix is equal to one, and we can write
\begin{eqnarray}\label{1446}
   \|{b}(\varepsilon)-{b}(0)\|\approx \varepsilon\left(\| \tilde{r}_{2n-1}\|+\| \tilde{R}_{2n-1}\|\right)\|{b}(0)\|.
\end{eqnarray}
\begin{exm}
Let us calculate the latter distance for the case of $l_2$ norm. 
\\By definition
\begin{eqnarray}\label{1330}
    \|R\|_2=\sup_{x\in \mathbbm{R}^{2n-1}\ \{0\}}\frac{\|Rx\|_2}{\|x\|_2},\quad \|x\|_2=\sqrt{\sum\limits_{i=1}^{2n-1}x_i^2}\geq \frac{1}{\sqrt{2n-1}}\sum\limits_{i=1}^{2n-1}|x_i|,
\end{eqnarray}
hold. 
We can write
\begin{eqnarray}
   \|\tilde{r}\|_2=\sqrt{\sum\limits_{i=1}^{2n-1}|\tilde{r}_i|^2}=1.
\end{eqnarray}
Let us introduce a constant $\gamma_0>0$ such that  vector $R_i(x)\in \mathbbm{R}^{2n-1\times1}$ is bounded by norm as $\|R_i(x)\|_2\leq \gamma_0$.
The matrix ${R}_{2n-1}=[R_1,R_2,\dots,R_{2n-1}]\in \mathbbm{R}^{2n-1\times 2n-1}$ holds. The constant is
\begin{eqnarray}
   \gamma_0\leq \sqrt{2n-1}R_{max},\quad R_{max}=\sqrt{(\max\limits_{j}(R_i(x))_j)^2},\quad \forall i\in\overline{1,2n-1}.
\end{eqnarray}
Then we can write
\begin{eqnarray}
    \|Rx\|_2=\Bigg\|\sum\limits_{i=1}^{2n-1}x_i R_i\Bigg\|_2\leq \sum\limits_{i=1}^{2n-1}|x_i|\|R_i\|_2\leq \gamma_0 \sum\limits_{i=1}^{2n-1}|x_i|.
\end{eqnarray}
Substituting it in \eqref{1330}, we get
\begin{eqnarray}\label{1331}
    \|R\|_2\leq \gamma_0 \sqrt{2n-1}\leq (2n-1)R_{max}.
\end{eqnarray}
Then
\begin{eqnarray}\label{1331_10}
    \|\tilde{R}\|_2\leq \sqrt{2n-1}R_{max}.
\end{eqnarray}
The $l_2$ norm of \eqref{1615} is
\begin{eqnarray}
    \|{b}_{2n-1}(0,t,\Delta)\|_2=\frac{2\Delta}{\sqrt{2n-1}}\sqrt{\sum\limits_{k=1}^{2n-2} \left(\sum\limits_{j=0}^{n-1}2^j\sin{((j+1) kt)}\right)^2}
\end{eqnarray}
We can upper bound it as
\begin{eqnarray}
    \|{b}_{2n-1}(0,t,\Delta)\|_2\leq \frac{4(n-1)(2^n-1)^2\Delta}{\sqrt{2n-1}}.
\end{eqnarray}
Then \eqref{1446} can be bounded by
\begin{eqnarray}\label{1459}
   \|{b}(\varepsilon,t,\Delta)-{b}(0,t,\Delta)\|_2\leq  4\varepsilon\Delta\left(1+\sqrt{2n-1}R_{max}\right)\frac{(n-1)(2^n-1)^2}{\sqrt{2n-1}}.
\end{eqnarray}
Then, for example, one can select 
\begin{eqnarray}
    \varepsilon\approx (4\Delta n(n-1)(2^n-1)^2)^{-1}
\end{eqnarray}
and \eqref{1459} tends to zero while $n\rightarrow\infty$.
\end{exm}
\subsubsection{Eigenvalues Forming Equidistant Sets}
\par
\begin{figure} [t]
    \begin{center} 
        \includegraphics[width=8.5cm]{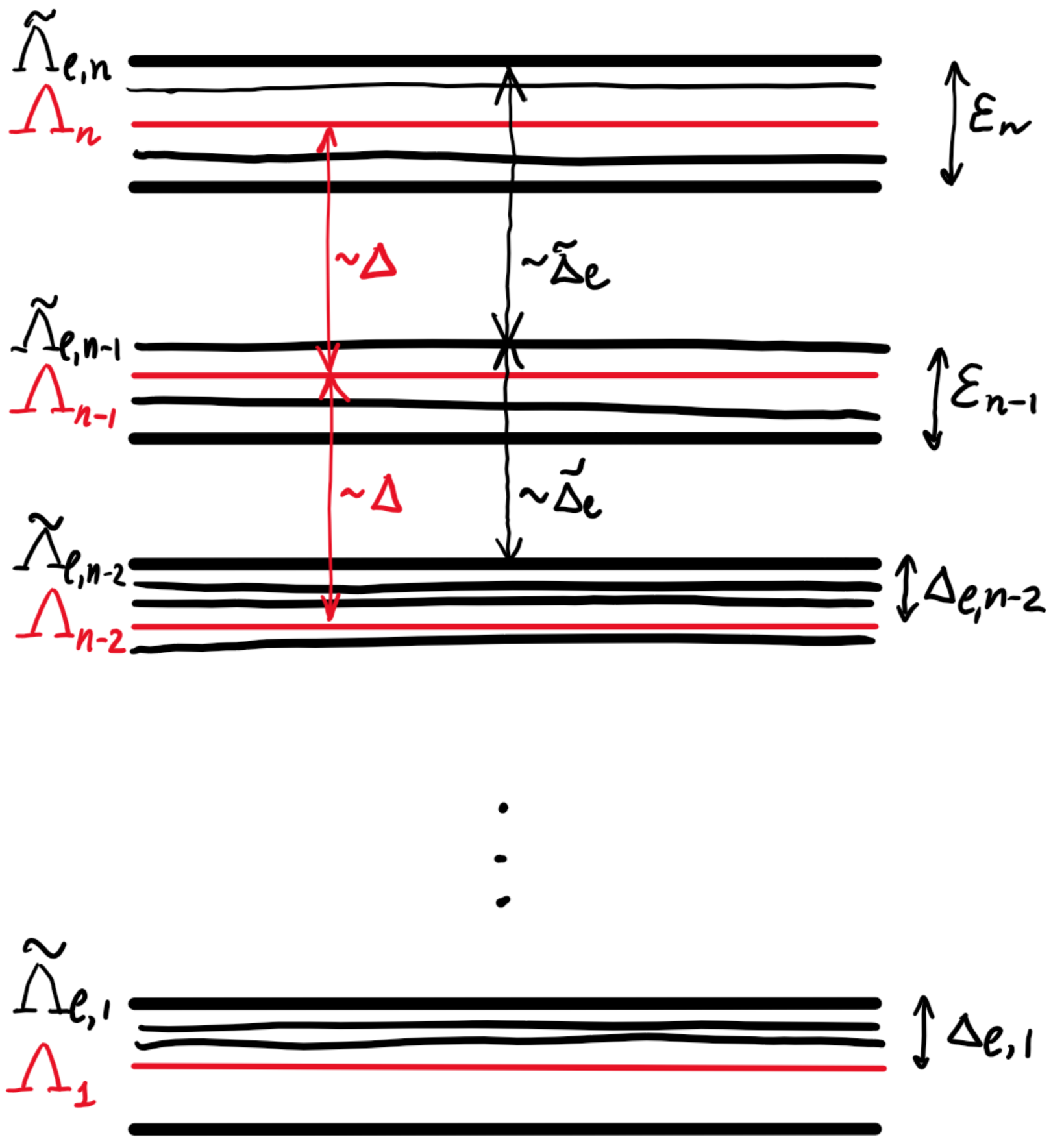}
        \end{center} 
         \caption{Equidistant shifted clusters of eigenvalues. \label{fig.2} }
    \end{figure}

If we have $k$ realizations of the Hamiltonian, the sets of eigenvalues $(\{\lambda_i\}_{i=1}^n)_k$ can be considered perturbed from each other.
Let us assume that we can sort all eigenvalues from $k$ realizations into $n$ equidistant sets (see Fig.~\ref{fig.2}) with median values denoted as ${\Lambda}_i$, $i=\overline{1,n}$. The distance between the median values of every two neighboring clusters is
\begin{eqnarray}\label{1458_1}\mu_{(i,i+1)}\equiv 
|{\Lambda}_{i}-{\Lambda}_{i+1}|\approx\Delta, \quad \forall i\in\overline{1,n}.
\end{eqnarray}
We demand the width of every set to be $\epsilon_{i}<<\Delta$, $\forall i\in\overline{1,n}$.
\par Let us consider the median eigenvalues. For the set of equidistant ${\Lambda}_i$, $i=\overline{1,n}$ the solution of the problem \eqref{72} is ${b}_{2n-1}(\tau,\Delta)$ and is given by \eqref{1615}.
\par We sort the eigenvalues into equidistant groups in such a way that from every set, only one eigenvalue is picked. The eigenvalues in every group are denoted as $\tilde{\Lambda}_{l,i}$, $\forall l\in\overline{1,k}$, $i\in \overline{1,n}$, where the first index is the number of the group and the second is the number of eigenvalues in it. The distance between any eigenvalue in one set and its median is
\begin{eqnarray}\label{1146_1}
    |\tilde{\Lambda}_{l,i}-{\Lambda}_{i}|=\Delta_{l,i}< \epsilon_i.
\end{eqnarray}
Then the collection of eigenvalues $\{\tilde{\Lambda}_{l,i}\}_{i=1}^{n}$ are slightly shifted from the centers of the sets, but not more than the width $\epsilon_i$ according to \eqref{1146_1}.
\par First, we consider the case when we picked the shifted from the median collection of eigenvalues in such a way that the new collection $\{\tilde{\Lambda}_{l,i}\}_{i=1}^{n}$ is equidistant too (see Fig.~\ref{fig.2}). That means that
\begin{eqnarray}\label{1146_11}
    |\tilde{\Lambda}_{l,i}-\tilde{\Lambda}_{l,i-1}|=\tilde{\Delta}_l.
\end{eqnarray}
For these $n$ eigenvalues, the solution  of the reduced problem \eqref{72} is ${b}_{2n-1}(\tau,\tilde{\Delta}_l)$ and is given by \eqref{1615}.
If $\Delta=\tilde{\Delta}$ the solutions from the median set and from the shifted set are coincident. In the real case these values can be slightly different. The shift is
\begin{eqnarray}
   \Delta+ \Delta_{l,i}- \Delta_{l,i-1}=\tilde{\Delta}_l,
\end{eqnarray}
where we considered the case when $\tilde{\Lambda}_{l,i}>{\Lambda}_{l}$.
Hence, we can write
\begin{eqnarray}
    {b}_{2n-1}(\tau,\tilde{\Delta}_l)= {b}_{2n-1}(\tau,\Delta+ \Delta_{l,i}- \Delta_{l,i-1})=
    {b}_{2n-1}(\tau,\Delta)+ {b}_{2n-1}(\tau,\Delta_{l,i})- {b}_{2n-1}(t,\Delta_{l,i-1}).
\end{eqnarray}
Using different collections of equidistant eigenvalues we can get a series of estimates of ${b}_{2n-1}(\tau,\Delta)$:
\begin{eqnarray}
    {b}_{2n-1}(\tau,\Delta)= {b}_{2n-1}(\tau,\tilde{\Delta}_l)- {b}_{2n-1}(\tau,\Delta_{l,i})+{b}_{2n-1}(t,\Delta_{l,i-1}).
\end{eqnarray}
\par  However, if we sort the real data, we will see that the eigenvalues are slightly not equidistant, corresponding to the case of perturbed equidistant eigenvalues we considered in the previous subsection. Then, one has to do the same analysis, taking into account the amount of perturbation from the equidistant positions.

\section{Phase Shift Rule for an Ill-Posed Problem}\label{sec_5}
\par 
In this section, we solve the problem \eqref{72}
being ill-posed by Hadamard. As we mentioned, it can happen for multiple reasons. First, the eigenvalues of the Hamiltonian can be close to each other, such that different $\mu_{(k,p)}$ would be equal. This causes singularity in the matrix $E_m$ (further, we omit the $m$ index, assuming all matrices are of size $m\times m$). Secondly, we solve \eqref{72} for the case when the operators $E$ and the functions $\mu$ are not known precisely but one knows their approximations $\hat{E}_l$ and $\hat{\mu}_l$ instead.
Here index $l\in \overline{1,L}$, $L>0$ denotes the realization number. The approximates $\hat{E}_l$ and $\hat{\mu}_l$ are defined on a probability space
$(\Omega,\mathcal{A},{P})$ and are close to $E$ and $\mu$ in some probabilistic sense. Here,  $\hat{\mu}_l\in V$ and
the operator $\hat{E}_l$ is continuous  $\forall\omega\in\Omega$.
\par Since $\hat{b}(\vec{\phi})=\hat{E}_l^{-1}(\vec{\phi})\hat{\mu}_l$ is unstable with respect to fluctuations in the empirical data, it cannot be utilized as an approximation of $b(\vec{\phi})$. To be more precise, slight variations in the values of $\hat{\mu}_l$ from $\mu$ have the potential to result in significant variations in $\hat{b}$. This implies that the inverse operator $\hat{E}_l^{-1}$ may not be continuous and the problem is ill-posed.
\par In our specific case, ${\mu}$ is a $(m\times 1)$ vector. If ${E}$ is an $m\times m$ matrix and $\det{{E}}\neq 0$ (or $\rank{({E})}=m$) then ${E}^{-1}$ exists. However, the problem can still be ill-posed. One can define an orthogonal transformation
$b=Vb^{\star}$ and $\mu=V\mu^{\star}$ such that $E$ will be represented in a diagonal form $(l_1,\dots,l_m)$, where $\{l_i\}_{i=1}^{m}$ are the eigenvalues of $E$.  When some differences $\mu_{(k,p)}$ between the eigenvalues of the Hamiltonian  are equal, the $\rank{(E)}=r<m$ and then the $m-r$ eigenvalues $l_i$ of the matrix $E$ are zero. Then the matrix is not invertible. Let $l_i=0$, $i=\overline{1,r}$  and $l_i\neq 0$ for $i\in\overline{r+1,m}$. 
For a given approximations $\hat{E}_l$ and $\hat{\mu}_l$ such that 
 \begin{eqnarray}
   && \|\hat{E}_l-E\|\leq \varepsilon,\quad \varepsilon>0,\\\nonumber &&\|\hat{\mu}_l-\mu\|\leq \delta,\quad \delta>0,
\end{eqnarray} the eigenvalues $\tilde{l}_i$, $i\in \overline{r+1,m}$ of $\hat{E}_l$ may be close to zero for a sufficiently small $\varepsilon$. Then $\tilde{b}_i^{\star}=\tilde{\lambda}_i^{\star}/\tilde{l}_i$ may be large for a small perturbation of $\hat{E}_l$ and $\hat{\mu}_l$. This implies that the solution of the system of linear equations \eqref{72} is unstable.
\par In this case, we use the regularization technique introduced by Tikhonov and Arsenin (1977) \cite{tikhonov1977solutions} that entails the stabilization of solutions by limiting the set of feasible solutions $\mathcal{D}\in U$ to a compact set $\mathcal{D}^{\star}$, due to the subsequent lemma:
\begin{lemma}
The inverse operator $E^{-1}$ is continuous on the set ${N}^{\star}=E\mathcal{D}^{\star}$ if the
continuous one-to-one operator $E$ is defined on the compact $\mathcal{D}^{\star}\in \mathcal{D}\subseteq U$.
\end{lemma}
The reduction of solutions is provided by the stabilizing functional, which is defined on $\mathcal{D}$. One can notice that the
regularization method is similar to the Lagrange method in the sense that we are looking for a solution $\hat{b}$ that minimizes a functional $\Omega(\hat{b}):\|\hat{E}_l \hat{b}-\hat{\mu}_l\|\leq \varepsilon$, $\varepsilon>0$.
\par In this paper, to find the solution of \eqref{72} we propose to use the extension of the regularization method from a
deterministic operator equation to the case of stochastic ill-posed problems. The
function that minimizes the functional
\begin{eqnarray}\label{73}
R_{\gamma}(\hat{\mu}_l,b)= \|\hat{E}_l b-\hat{\mu}_l\|^2_V+\gamma \Omega(b),
\end{eqnarray}
in a set $\mathcal{D}$ of functions $b\in U$ is taken as an approximate solution of \eqref{72}. The parameter $\gamma >0$ is called the regularization parameter and $\Omega(b)$ is a stabilizing functional that satisfies the following conditions:
\begin{itemize}
 \item $\Omega(b)$ is defined on the set $\mathcal{D}$.
\item $\Omega(b)$ assumes real nonnegative values and is lower semi-continuous on $\mathcal{D}$.
 \item All set $M_c=\{b:\Omega(b)\leq c\}$ are compact in $U$.
\end{itemize}
Further Theorems~\ref{th2} and \ref{th3}~\cite{vapnik1979nonparametric,vapnik2006estimation} provide the theoretical background of the statistical
regularization method for the case of an accurately given operator $E$, and Theorem~\ref{th4}~\cite{stefanyuk1986estimating} for the case of an inaccurately
given operator $E$.
\begin{theorem}\label{th2}
    If, for each $l$, a positive $\gamma=\gamma(l)$ is chosen such that $\gamma\rightarrow 0$ as
$l\rightarrow \infty$, then for any positive $\alpha$ and  $\beta$ there will be a number $N=N(\alpha,\beta)$ such
that, for all $l>N$, the elements $\hat{b}^{\gamma}(x)$ that minimize the functional \eqref{73} satisfy the
inequality
\begin{eqnarray}
    P\{\rho_U(\hat{b}^{\gamma},b)>\alpha\}\leq P\{\rho_V^2(\hat{\mu}_l,\mu)>\beta\gamma\},
\end{eqnarray}
where $b$ is the precise solution of \eqref{72} with the right-hand side $\mu$, and 
$\rho(f,g)=\|f-g\|$.
\end{theorem}
For our concrete case, all spaces are Hilbert ones. The following theorems state:
\begin{theorem}\label{th3}
Let $U$ be a Hilbert space, $E$ be a linear operator, and $\Omega(b)=\|b\|^2_U$. Then, $\forall\varepsilon$, there exists a number $l(\varepsilon)$ such that $\forall k>k(\varepsilon)$ the inequality 
\begin{eqnarray}
    P\{\|\hat{b}^{\gamma}-b\|^2_U>\varepsilon\}\leq 2P\{\rho_V^2(\hat{\mu}_l,\mu)>(\varepsilon/2) \gamma\},
\end{eqnarray}
holds.
\end{theorem}
\begin{theorem}\label{th4}
Let $U$ and $V$ be normed spaces. For any $\varepsilon>0$ and any constants $c_1,c_2>0$, there exists a number $\gamma_0>0$ such that  $\forall\gamma\geq \gamma_0$, 
\begin{eqnarray}
    P\{\omega:\|\hat{b}^{\gamma}-b\|_U>\varepsilon\}\leq P\{\omega:\frac{\|\hat{\mu}_l-\mu\|_V}{\sqrt{\gamma}}>c_1\}+ P\{\omega:\frac{\|\hat{E}_l-E\|}{\sqrt{\gamma}}>c_2\},
\end{eqnarray}
where 
\begin{eqnarray}
    \|\hat{E}_l-E\|=\sup\limits_{g\in\mathcal{D}}{\frac{\|\hat{E}_l b-E b\|_V}{\sqrt{\Omega(b)}}}.
\end{eqnarray}
\end{theorem}
These theorems imply that the minimization of \eqref{73} is a stable problem, i.e.
close functions $\hat{\mu}_l$ and $\mu$ (and close operators $\hat{E}_l$ and $E$) correspond to close (in
probabilistic sense) regularized solutions $\hat{b}^{\gamma}$ and $b$ that minimize the functionals
$  R_{\gamma}(\hat{\mu}_l,b)$ and $  R_{\gamma}(\mu,b)$, respectively.
\par For the Hilbert spaces $U$ and $V$, the solution of \eqref{72} with $\Omega(b)=\|b\|^2_U$ has a simple form
\begin{eqnarray}\label{1556}
    \hat{b}^{\gamma}=(\gamma I+\hat{E}_l^{\dagger} \hat{E}_l)^{-1}\hat{E}_l^{\dagger}\hat{\mu}_l,
\end{eqnarray}
where $I$ is a unit operator. 
\par The stability of the
approximation $\hat{b}^{\gamma}$ to $b$ 
 is ensured by an appropriate choice of $\gamma$. 
For selecting the regularization parameter, see \cite{morozov1984regular,engl1988posteriori,vapnik1992rate}. For example, the mismatch method ~\cite{morozov1984regular}
determines $\gamma$ from the equality
\begin{eqnarray}\label{1042}
   && \|\hat{E}_l \hat{b}^{\gamma}-\hat{\mu}_l\|_V=\varepsilon(l)+\eta(l,b),\\\nonumber
    &&\|\hat{\mu}_l-\mu\|_V\leq \varepsilon(l),\quad \|\hat{E}_l b-Eb\|_V\leq \eta(l,b),
\end{eqnarray}
where $\varepsilon(l)$ and $\eta(l,b)$ are known estimates of the data error. The stochastic analog of the mismatch
method is the discrepancy method~\cite{cheng2015regularization,lu2010generalized}. If the operator is defined precisely ($\eta(l,b)=0$), then the choice of $\gamma$ from
\eqref{1042} provides a rate of convergence of the regularized estimate $\hat{b}^{\gamma}$
to $b$ that is no
better than $O(\varepsilon^{1/2})$ (see \cite{engl1988posteriori}).
\subsection{Minimization of the Square Norm}
\par Ones we know the approximate solution of \eqref{72} defined by \eqref{1556}, we can solve the  minimization problem  \eqref{1255} to minimize the variance \eqref{1219}.
Using the form of the regularized solution \eqref{1556} of \eqref{72}, we can write
\begin{eqnarray}
    \hat{b}^{\gamma}_x(\vec{\phi})=\sum\limits_{j,i=1}^{m}(\gamma I+\hat{E}^{\dagger}(\vec{\phi}) \hat{E}(\vec{\phi}))_{xj}^{-1}\hat{E}_{ji}^{\dagger}(\vec{\phi})\hat{\mu}_i,
\end{eqnarray} 
where we omit the index $l$ meaning we treat one experimental realization of $E$ and $\mu$. 
The derivative is
\begin{eqnarray}
  \left( \frac{\partial \hat{b}^{\gamma}(\phi)}{\partial \phi_y}\right)_x=\sum\limits_{j,s=1}^{m}\left(\left(\frac{\partial (\gamma I+\hat{E}^{\dagger} \hat{E})^{-1}}{\partial \phi_y}\right)_{xj}\hat{E}_{js}^{\dagger}+(\gamma I+\hat{E}^{\dagger} \hat{E})_{xj}^{-1}  \left( \frac{\partial \hat{E}^{\dagger}}{\partial \phi_y}\right)_{js}\right)\hat{\mu}_s,
\end{eqnarray}
where we use the short notation omitting $\phi$ dependence. 
Using $\frac{\partial Y^{-1}(x) }{\partial x}=-Y^{-1}\frac{\partial Y(x) }{\partial x}Y^{-1}$, we get
\begin{eqnarray}\label{1047}
\left(\frac{\partial (\gamma I+\hat{E}^{\dagger} \hat{E}}{\partial \phi_y}\right)_{xj}=-\sum\limits_{l,p=1}^{m}(\gamma I+\hat{E}^{\dagger} \hat{E})_{x l}^{-1}\left(\frac{\partial (\gamma I+\hat{E}^{\dagger} \hat{E})}{\partial \phi_y}\right)_{lp}
(\gamma I+\hat{E}^{\dagger} \hat{E})_{pj}^{-1}.
\end{eqnarray}
The derivative is
\begin{eqnarray}
    \left(\frac{\partial (\gamma I+\hat{E}^{\dagger} \hat{E})}{\partial \phi_y}\right)_{lp}=\sum\limits_{v=1}^{m}\left(\frac{\partial E^{\dagger} }{\partial \phi_y}\right)_{lv}E_{vp}\delta_{l,y}+E^{\dagger}_{lv}\left(\frac{\partial E }{\partial \phi_y}\right)_{vp}\delta_{p,y},
\end{eqnarray}
where we used the fact that the derivatives on the right hand side are non zero only in one raw or column. 
\par Finally, the expression \eqref{1255} for the regularised solution can be written  as follows
\begin{eqnarray}\label{1443_2}
&&\sum\limits_{x=1}^{m}\sum\limits_{j,i=1}^{m}(\gamma I+\hat{E}^{\dagger} \hat{E})_{xj}^{-1}\hat{E}_{ji}^{\dagger}\lambda_i \sum\limits_{l=1}^{m}(\gamma I+\hat{E}^{\dagger} \hat{E})_{xl}^{-1}\sum\limits_{s=1}^{m}\Bigg(\left( \frac{\partial \hat{E} ^{\dagger}}{\partial \phi_y}\right)_{ls}\delta_{l,y} \\\nonumber
&-&\sum\limits_{p,v,t=1}^{m}\left(\left(\frac{\partial \hat{E}^{\dagger} }{\partial \phi_y}\right)_{lv}\hat{E}_{vp}\delta_{l,y}+\hat{E}^{\dagger}_{lv}\left(\frac{\partial \hat{E} }{\partial \phi_y}\right)_{vp}\delta_{p,y}\right)
(\gamma I+\hat{E}^{\dagger} \hat{E})_{pt}^{-1}\hat{E}_{ts}^{\dagger} \Bigg)\lambda_s=\mathbf{0}.
\end{eqnarray}
Solving this system of equation with respect to all $\phi_y$, $y\in\overline{1,m}$, one can find the optimal $\vec{\phi}$ minimizing the variance \eqref{1219}.
\par A possible solution is, when
\begin{eqnarray*}
   \left( \frac{\partial E ^{\dagger}}{\partial \phi_y}\right)_{ls}\delta_{l,y}=\sum\limits_{p,v,t=1}^{m}\left(\left(\frac{\partial E^{\dagger} }{\partial \phi_y}\right)_{lv}E_{vp}\delta_{l,y}+E^{\dagger}_{lv}\left(\frac{\partial E }{\partial \phi_y}\right)_{vp}\delta_{p,y}\right)
(\gamma I+E^{\dagger} E)_{pt}^{-1}E_{ts}^{\dagger},\quad \forall l,s=\overline{1,m}, 
\end{eqnarray*}
holds.
Then 
\begin{eqnarray*}
   &&\left( \frac{\partial E ^{\dagger}}{\partial \phi_y}\right)_{ys}=\sum\limits_{p,v=1}^{m}\left(\frac{\partial E^{\dagger} }{\partial \phi_y}\right)_{yv}E_{vp}\sum\limits_{t=1}^{m}(\gamma I+E^{\dagger} E)_{pt}^{-1}E_{ts}^{\dagger}\\
   &+&\sum\limits_{v=1}^{m}E^{\dagger}_{yv}\left(\frac{\partial E }{\partial \phi_y}\right)_{vy}
\sum\limits_{t=1}^{m}(\gamma I+E^{\dagger} E)_{yt}^{-1}E_{ts}^{\dagger},\quad \forall s=\overline{1,m}\quad \text{and}\quad l=y;\\\nonumber
&&\sum\limits_{v=1}^{m}E^{\dagger}_{lv}\left(\frac{\partial E }{\partial \phi_y}\right)_{vy}\sum\limits_{t=1}^{m}
(\gamma I+E^{\dagger} E)_{yt}^{-1}E_{ts}^{\dagger}=\mathbf{0},\quad \forall s=\overline{1,m}\quad \text{and}\quad l\neq y. 
\end{eqnarray*}

\section{Discussion and Conclusion}\label{sec_7}
\par We propose the phase shift rule, with the optimal parameter selection dependent on the Hamiltonian eigenvalue structure. Our method is suitable for big Hamiltonian systems with known eigenvalues. Dependent on the distance between the eigenvalues, the problem can be well- or ill-posed by Hadamard, which makes it non-trivial for optimization in the case when some distances are close to each other.
\par In the case of a well-posed problem, an explicit solution is proposed, and the recipe for finding the optimal phases is provided. For the ill-posed problem arising, for example, when the eigenvalues of the Hamiltonian are close to each other and the distances between them can coincide, we find the explicit solution as well. We show that it is unique and that the phases must be picked equidistantly. We observe the realistic case of slightly perturbed equidistant eigenvalues arising in practice and the case of equidistant clusters formed by the different realizations of the Hamiltonian.
We provide the regularized solution for the ill-posed problem that does not have a particular eigenvalue structure, as well as the method of optimum phase shift selection.
\par In addition to a full reconstruction of the derivative, the
presented approach offers parameter-shift rules for
derivatives of arbitrary order and any linear combination of them.

\section{Acknowledgments}
L.M. was  supported by the Netherlands Organisation for Scientific Research (NWO/OCW), as part of the Quantum Software Consortium program (project number 024.003.037 / 3368).
This work has received support from the European Union's Horizon Europe research and innovation programme through the ERC StG FINE-TEA-SQUAD (Grant No. 101040729). This work is supported by the Dutch National Growth Fund (NGF), as part of the Quantum Delta NL programme.

Funded by the European Union. Views and opinions expressed are however those of the author(s) only and do not necessarily reflect those of the European Union or the European Commission. Neither the European Union nor the granting authority can be held responsible for them.

\bibliographystyle{unsrt}
\bibliography{Bib}

\appendix
\section{Hamiltonian with Finite Number of Eigenvalues}\label{app_0}
Suppose a Hamiltonian $H$ has $n$ distinct eigenvalues $\vec{\lambda}=\{\lambda_1,\lambda_2,\dots,\lambda_n\}$. Then, one can write the characteristic polynomial of its matrix as:
\begin{equation}
    p(H)=\prod_{i=1}^n(H-\lambda_i \mathbb{I})=0.
\end{equation}
This means that there exist only $n-1$ non-trivial powers of $H$, since one can write:
\begin{equation}
    H^n=-\sum_{i=1}^{n-1}(-1)^{i}S_{i}(\vec{\lambda})H^{n-i},
\end{equation}
where the term $S_i(\vec{\lambda})$ is the symmetric polynomial of degree $i$ that contains all the possible combinations of elements of $\vec{\lambda}$ (of degree $1$). For example, if $n=3$ and $i=2$: 

\begin{equation}
    S_2=\lambda_1\lambda_2+\lambda_1\lambda_3+\lambda_2\lambda_3.
\end{equation}
The fact that there are only finite number of powers means that also any function of $H$ can be written as follows
\begin{equation}
\label{seriesham}
    f(H)=\sum_{i=0}^{n-1} c_i H^{i}:=\braket{H}{c}.
\end{equation}
To determine the form of the vector $c_i$, one can project \eqref{seriesham} onto the eigenstates of $H$. Schematically we can write it:
\begin{equation}
   \ket{f(\vec\lambda)}=\Lambda(\vec{\lambda})\ket{c} \Rightarrow \ket{c}=\Lambda^{-1}(\vec{\lambda})\ket{f(\vec \lambda)}.
\end{equation}
In the above expression the vector $\ket{f(\vec \lambda)}$ contains the $f(\lambda_i), \forall \lambda_i\in\vec \lambda$ and $\Lambda$ is the $n
\times n$ Vandermonde matrix containing the $\vec \lambda$. The explicit form of this square matrix is:
\begin{equation}
    \big(\Lambda^{-1}\big)_{ij}=(-1)^{i+1}\frac{S_{n-1-i}(\vec \lambda_j)}{\prod_{k\in \vec \lambda_j}(k-\lambda_j)}
\end{equation}
This is the element that belongs to i-th row and j-th column. The $\vec \lambda_j$ denotes the set of $\vec \lambda$ that does not include the single element $\lambda_j$.
Then the coefficients  in \eqref{seriesham} are the following
\begin{gather}
    c_i(f)=(-1)^{i+1}\sum_{j=1}^n \frac{S_{n-1-i}(\vec\lambda_j)}{\prod_{k\in \vec\lambda_j}(k-\lambda_j)}f(\lambda_j).
    \end{gather}
\par Let us consider the function  $f(H)=\exp(-\mathbbm{i} H t)$. Based on the previous results the coefficients  in the decomposition \eqref{seriesham} are the following
\begin{gather}
    c_p(e)=(-1)^{p+1}\sum_{j=1}^n \frac{S_{n-1-p}(\vec\lambda_j)}{\prod_{k\in \vec\lambda_j}(k-\lambda_j)}\exp(-\mathbbm{i}\lambda_jt).
\end{gather}
This implies that the coefficients of the derivative of $f(H)$ with respect to $t$ can be written in a closed form as follows

\begin{equation}
    \tilde{c}_p(e)=(-1)^{p+1}\sum_{j=1}^n\frac{S_{n-1-p}(\vec\lambda_j)\lambda_j}{\prod_{k\in \vec\lambda_j}(k-\lambda_j)}\exp(-\mathbbm{i}(\lambda_jt+\frac{\pi}{2})).
\end{equation}

 \section{Not Equidistant Phases}\label{ap_1}
\par We want to solve the system
\begin{eqnarray}\label{1456_4}
   \phi_{i}-\phi_{j}=\frac{2\pi }{(2n-1)\Delta} t_{ij},\quad t_{ij}\in Z,\quad \forall i,j=[1,{2n-1}],\quad  i<j.
\end{eqnarray}
Let us investigate the example of $n=3$, then $m=5$. Let $\phi_5=0$, holds, and we can write:
\begin{eqnarray}
   && \phi_1=\frac{t}{\Delta}t_{15},\quad 
    \phi_2=\frac{t}{\Delta}(t_{15}+c_1\epsilon),\quad 
     \phi_3=\frac{t}{\Delta}(t_{15}+c_2\epsilon),\quad 
      \phi_4=\frac{t}{\Delta}(t_{15}+c_3\epsilon),\\\nonumber
      && 
    \phi_2=\frac{t}{\Delta}t_{25},\quad 
     \phi_3=\frac{t}{\Delta}(t_{25}+b_1\epsilon),\quad 
      \phi_4=\frac{t}{\Delta}(t_{25}+b_2\epsilon),
      \\\nonumber
      && 
     \phi_3=\frac{t}{\Delta}t_{35},\quad 
      \phi_4=\frac{t}{\Delta}(t_{35}+a_1\epsilon),
      \\\nonumber
      && 
     \phi_4=\frac{t}{\Delta}t_{45}. 
\end{eqnarray}
where we denoted $t_{1,2}=c_1\epsilon$, $t_{1,3}=c_2\epsilon$, $t_{1,4}=c_3\epsilon$,
     $t_{2,3}=b_1\epsilon$, $t_{2,4}=b_2\epsilon$, $t_{3,4}=a_1\epsilon$.
The constants can be easily found
\begin{eqnarray}
   && c_1 = \frac{1}{\epsilon} (t_{25} - t_{15}),\quad 
c_2 =\frac{1}{\epsilon}  (t_{35} - t_{15}),\quad 
c_3 = \frac{1}{\epsilon} (t_{45} - t_{15}),\\\nonumber
&&b_1 =\frac{1}{\epsilon}  (t_{35}- t_{25}),\quad 
b_2 =\frac{1}{\epsilon}  (t_{45} - t_{25}),\quad 
a_1 = \frac{1}{\epsilon}  (t_{45} - t_{35}).
\end{eqnarray}
Selecting any $t_{15}<t_{25}<t_{35}<t_{45}$, one get solutions:
\begin{eqnarray}
     && \phi_1=\frac{t}{\Delta}t_{15},\quad 
    \phi_2=\frac{t}{\Delta}t_{25},\quad 
     \phi_3=\frac{t}{\Delta}t_{35},\quad 
     \phi_4=\frac{t}{\Delta}t_{45},\quad 
     \phi_5=0.
\end{eqnarray}
One can see that the equation \eqref{1456} is solvable. The solution is
\begin{eqnarray}
    \phi_i=\frac{t}{\Delta}t_{i,2n-1}, \quad \phi_{2n-1}=0,\quad \forall i\in[1,2n-2].
\end{eqnarray}
However, due to periodicity of the complex exponent, we get singular matrix $E_{2n-1}$ in all cases except of equidistant phases one. The latter example can be easily distributed to any value of $n$.

 \section{Not Equidistant Eigenvalues}\label{ap_2}

Let $\{\lambda_i\}_{i=1}^n$ be equidistant eigenvalues. We consider another set $\{\tilde{\lambda}_i\}_{i=1}^n$ such that
\begin{eqnarray}
    \tilde{\lambda}_i-\lambda_i=\Delta_i,\quad \Delta_i>0,\quad \forall i\in\overline{1,n}.
\end{eqnarray}
Then the distance between the new set of eigenvalues is
\begin{eqnarray}
   \tilde{ \mu}_{(i,j)}=\tilde{\lambda}_i-\tilde{\lambda}_j=\lambda_i-\lambda_j+\Delta_i-\Delta_j={ \mu}_{(i,j)}+\Delta_{ij}.
\end{eqnarray}
where we used the notation $\Delta_{ij}\equiv \Delta_i-\Delta_j$.
Then the matrix $\hat{E}_k$ for these eigenvalues can be written as
\begin{eqnarray}\!\!\!\hat{E}_k=\!\!
\begin{bmatrix}
1 & 1 & \dots& 1  \\
e^{\mathbbm{i}{ \mu}_{(1,2)}\phi_1}e^{\mathbbm{i}\Delta_{1,2}\phi_1} &e^{\mathbbm{i}{ \mu}_{(1,2)}\phi_2}e^{\mathbbm{i}\Delta_{1,2}\phi_2} & \dots& e^{\mathbbm{i}{ \mu}_{(1,2)}\phi_m}e^{\mathbbm{i}\Delta_{1,2}\phi_m} \\
e^{-\mathbbm{i}{ \mu}_{(1,2)}\phi_1}e^{-\mathbbm{i}\Delta_{1,2}\phi_1} &e^{-\mathbbm{i}{ \mu}_{(1,2)}\phi_2}e^{-\mathbbm{i}\Delta_{1,2}\phi_2}& \cdot& e^{-\mathbbm{i}{ \mu}_{(1,2)}\phi_m}e^{-\mathbbm{i}\Delta_{1,2}\phi_m}  \\
\vdots &\vdots&   \ddots& \vdots\\
e^{-\mathbbm{i}{ \mu}_{(n,n-1)}\phi_1}e^{-\mathbbm{i}\Delta_{n,n-1}\phi_1} &e^{-\mathbbm{i}{ \mu}_{(n,n-1)}\phi_2}e^{-\mathbbm{i}\Delta_{n,n-1}\phi_2} & \dots& e^{-\mathbbm{i}{ \mu}_{(n,n-1)}\phi_m}e^{-\mathbbm{i}\Delta_{n,n-1}\phi_m}
\end{bmatrix}.
\end{eqnarray}
We may represent it like a system with perturbation, basically $\hat{E}_k=E+\Omega$, where $\Omega$ represents the perturbation.
Let us use the small angle approximation 
\begin{eqnarray}
    \exp{(\mathbbm{i}\Delta_{i,j}\phi_k)}\approx 1+\mathbbm{i}\Delta_{i,j}\phi_k,\end{eqnarray}
    assuming  all $\Delta_{i,j}$ to be sufficiently small. Then we can write
\begin{eqnarray}
&&\hat{E}_k\approx E_m+\Omega,\\\nonumber
&&\Omega=\mathbbm{i}
\begin{bmatrix}
0 & 0 & \dots& 0  \\
e^{\mathbbm{i}{ \mu}_{(1,2)}\phi_1}\Delta_{1,2}\phi_1 &e^{\mathbbm{i}{ \mu}_{(1,2)}\phi_2}\Delta_{1,2}\phi_2 & \dots& e^{\mathbbm{i}{ \mu}_{(1,2)}\phi_m}\Delta_{1,2}\phi_m \\
-e^{-\mathbbm{i}{ \mu}_{(1,2)}\phi_1}\Delta_{1,2}\phi_1 &-e^{-\mathbbm{i}{ \mu}_{(1,2)}\phi_2}\Delta_{1,2}\phi_2& \cdot& -e^{-\mathbbm{i}{ \mu}_{(1,2)}\phi_m}\Delta_{1,2}\phi_m  \\
\vdots &\vdots&   \ddots& \vdots\\
-e^{-\mathbbm{i}{ \mu}_{(n,n-1)}\phi_1}\Delta_{n,n-1}\phi_1&-e^{-\mathbbm{i}{ \mu}_{(n,n-1)}\phi_2}\Delta_{n,n-1}\phi_2 & \dots& -e^{-\mathbbm{i}{ \mu}_{(n,n-1)}\phi_m}\Delta_{n,n-1}\phi_m
\end{bmatrix}
\end{eqnarray}
We can assume that $\Delta_{i,j}<\epsilon$, where $\epsilon\equiv \max\limits_{i,j}{(\Delta_{i,j})}$, $\forall i,j\in\overline{1,m}$. Then we can write
\begin{eqnarray}
\hat{E}_k&\approx&E_m+\epsilon R_m,\\ \nonumber
R_m&=&\mathbbm{i}
\begin{bmatrix}
0 & 0 & \dots& 0  \\
e^{\mathbbm{i}{ \mu}_{(1,2)}\phi_1}\phi_1 &e^{\mathbbm{i}{ \mu}_{(1,2)}\phi_2}\phi_2 & \dots& e^{\mathbbm{i}{ \mu}_{(1,2)}\phi_m}\phi_m \\
-e^{-\mathbbm{i}{ \mu}_{(1,2)}\phi_1}\phi_1 &-e^{-\mathbbm{i}{ \mu}_{(1,2)}\phi_2}\phi_2& \cdot& -e^{-\mathbbm{i}{ \mu}_{(1,2)}\phi_m}\phi_m  \\
\vdots &\vdots&   \ddots& \vdots\\
-e^{-\mathbbm{i}{ \mu}_{(n,n-1)}\phi_1}\phi_1&-e^{-\mathbbm{i}{ \mu}_{(n,n-1)}\phi_2}\phi_2 & \dots& -e^{-\mathbbm{i}{ \mu}_{(n,n-1)}\phi_m}\phi_m
\end{bmatrix}.
\end{eqnarray}
Let us remind, that on the right hand side of the initial  equation stands the vector ${\lambda}_m$ from the differences $\tilde{\mu}_{(k,l)}$, $k,l\in \overline{1, n}$. Then
\begin{eqnarray}
\hat{\lambda}_m=\vec{\lambda}_m+\mathbbm{i}\epsilon I_m.
\end{eqnarray}
The matrix $E_m$ is singular by construction. So we reduce it to $E_{2n-1}$ that is non-singular as we mentioned in the main text. The same we do with $R_{2n-1}$. Hence we define a shifted non-singular matrix, where we take intro account that the phases are defined in 
 \eqref{1234}, and write
\begin{eqnarray}
    \hat{E}_{2n-1}&\approx&E_{2n-1}+\epsilon R_{2n-1},\\ \nonumber
R_{2n-1}&=&\frac{\mathbbm{i}t}{\Delta}
\begin{bmatrix}
0 & 0 & \dots& 0  \\
e^{-\mathbbm{i}t} &2e^{-2\mathbbm{i}t}  & \dots& (2n-1)  \\
-e^{\mathbbm{i}t} &-2e^{2\mathbbm{i}t}& \cdot& - (2n-1)  \\
\vdots &\vdots&   \ddots& \vdots\\
-e^{\mathbbm{i}(n-1)t}&-2e^{\mathbbm{i}(n-1)t}& \dots& - (2n-1)
\end{bmatrix}.
\end{eqnarray}
that is a realistic way of introducing small perturbations. Then our perturbed problem can be written as
\begin{eqnarray}
    &&(\tilde{E}_{2n-1}+\epsilon \tilde{R}_{2n-1})\vec{b}(\epsilon,\vec{\phi})=  (\vec{\tilde{\lambda}}_{2n-1}+\epsilon \tilde{r}),\\\nonumber
    && \tilde{R}_{2n-1}\equiv \frac{ R_{2n-1}}{\sqrt{2n-1}} ,\quad \tilde{r}\equiv \frac{1}{\sqrt{2n-1}}\mathbbm{i}  I_{2n-1}.
\end{eqnarray}
\section{Deduction of the minima equation}\label{ap_3}
Substituting \eqref{1230} in \eqref{1255}, we get
\begin{eqnarray} \label{1507}&&\sum\limits_{x=1}^{m}\det(E(\vec{\phi}/\phi_x)) (\det E(\vec{\phi}))^{-2}\\\nonumber
&\times&\left(
\frac{\partial \det(E(\vec{\phi}/\phi_x))}{\partial \phi_y}
   -(\det E(\vec{\phi}))^{-1}\det(E(\vec{\phi}/\phi_x))
\frac{\partial (\det E(\vec{\phi}))}{\partial \phi_y}   
   \right)=\mathbf{0}.
\end{eqnarray}
Using the product rule, we can write
\begin{eqnarray}
    &&\frac{\partial (\det E(\vec{\phi}/\phi_x))}{\partial \phi_y}=\frac{\partial (\det(\vec{v}{(\phi_1)},\vec{v}(\phi_2),\dots,\mu,\dots,\vec{v}(\phi_m))}{\partial \phi_y}\\\nonumber
    &=&\det(\frac{\partial\vec{v}{(\phi_1)}}{\partial \phi_y},\vec{v}(\phi_2),\dots,\mu,\dots,\vec{v}(\phi_m))+
    \det(\vec{v}(\phi_1),\frac{\partial\vec{v}{(\phi_2)}}{\partial \phi_y},\dots,\mu,\dots,\vec{v}(\phi_m))\\
    &+&\dots+\det(\vec{v}(\phi_1),\vec{v}{(\phi_2)},\dots,\mu,\dots,\frac{\partial\vec{v}(\phi_m)}{\partial \phi_y}).
\end{eqnarray}
If $y\neq x$, then
\begin{eqnarray}
    &&\frac{\partial (\det E(\vec{\phi}/\phi_x))}{\partial \phi_y}=
    \det(\vec{v}(\phi_1),\dots,\frac{\partial\vec{v}{(\phi_y)}}{\partial \phi_y},\dots,\mu,\dots,\vec{v}(\phi_m))\equiv \det( E_y(\vec{\phi}/\phi_x)).
\end{eqnarray}
If $y= x$, then
\begin{eqnarray}
    &&\frac{\partial (\det E(\vec{\phi}/\phi_x))}{\partial \phi_x}=0
\end{eqnarray}
since the determinant of a matrix with two similar columns or a single zero column is equal to zero. We denote $\det( E_x(\vec{\phi}/\phi_x))=0$. Next the derivative
\begin{eqnarray}
   \frac{\partial (\det E(\vec{\phi}))}{\partial \phi_y}
    &=&\det(\vec{v}(\phi_1),\dots,\frac{\partial\vec{v}{(\phi_y)}}{\partial \phi_y},\dots,\vec{v}(\phi_m))\equiv \det( E_y(\vec{\phi})),
\end{eqnarray}
holds. 
We can rewrite \eqref{1507} as
\begin{eqnarray} \label{1508_1}\sum\limits_{x=1}^{m}\det(E(\vec{\phi}/\phi_x))
\det( E_y(\vec{\phi}/\phi_x))
   =(\det E(\vec{\phi}))^{-1}\det( E_y(\vec{\phi}))\sum\limits_{x=1}^{m}\left(\det(E(\vec{\phi}/\phi_x))\right)^2
.
\end{eqnarray}
Solving the latter system of equation with respect to all $\phi_y$, $y\in\overline{1,m}$, one can find the optimal $\vec{\phi}$ minimizing the variance \eqref{1219}.
Using $\det{AB}=\det{A}\det{B}$, we rewrite the latter expression as follows
\begin{eqnarray} \label{1509}\!\!\!\sum\limits_{x=1}^{m}\det(E(\vec{\phi}/\phi_x)\times E_y(\vec{\phi}/\phi_x))
   =\sum\limits_{x=1}^{m}\det (E^2(\vec{\phi}/\phi_x)\times E^{-1}(\vec{\phi})\times E_y(\vec{\phi})), \forall y\in\overline{1,m}
.
\end{eqnarray}
One of the straightforward solutions 
\begin{eqnarray}
   && E_y(\vec{\phi}/\phi_x)=E(\vec{\phi}/\phi_x)\times E^{-1}(\vec{\phi})\times E_y(\vec{\phi}),\quad \forall x,y\in\overline{1,m}, x\neq y,\\\nonumber
    &&0=E(\vec{\phi}/\phi_x)\times E^{-1}(\vec{\phi})\times E_x(\vec{\phi}),\quad  x= y
\end{eqnarray}
is not valid, since the matrices in the last line are not zero.

\end{document}